\documentclass[10pt,twocolumn,letterpaper]{article}

\usepackage{iccv}      

\usepackage{array}
\usepackage{comment}
\usepackage{tabularx}
\usepackage{xcolor}

\newcommand{\method}{GEOPARD}

\newcommand{\todo}[1]{{\color{red}#1}}

\definecolor{revolute}{rgb}{0.82,0.32,0.32}
\definecolor{prismatic}{rgb}{0.802, 0.5, 0.402}
\definecolor{candidate}{rgb}{0.5647058823529412, 0.8235294117647058, 0.9254901960784314}

\newcommand{\R}{\mathbb{R}}    
\newcommand{\mB}{\mathbf{B}}   

\newcommand{\ba}{\mathbf{a}}

\newcommand{\bc}{\mathbf{c}}

\newcommand{\bff}{\mathbf{f}}
\newcommand{\bg}{\mathbf{g}}
\newcommand{\bh}{\mathbf{h}}

\newcommand{\bm}{\mathbf{m}}

\newcommand{\bo}{\mathbf{o}}
\newcommand{\bp}{\mathbf{p}}
\newcommand{\bq}{\mathbf{q}}
\newcommand{\br}{\mathbf{r}}

\newcommand{\bv}{\mathbf{v}}

\newcommand{\mS}{\mathcal{S}}
\newcommand{\mP}{\mathcal{P}}

\def\R{{\cal R}}

\definecolor{iccvblue}{rgb}{0.21,0.49,0.74}
\usepackage[pagebackref,breaklinks,colorlinks,allcolors=iccvblue]{hyperref}
\usepackage{wrapfig}

\def\paperID{4539} 
\def\confName{ICCV}
\def\confYear{2025}

\title{\textcolor{teal}{\method}: \textcolor{teal}{Geo}metric \textcolor{teal}{P}retraining for 
\textcolor{teal}{Ar}ticulation Pre\textcolor{teal}{d}iction 
in 3D Shapes}

\author{Pradyumn Goyal$^1$~~~Dmitry Petrov$^1$~~~Sheldon Andrews$^2$~~~Yizhak Ben-Shabat$^3$
	\vspace{0.1cm}\\
	Hsueh-Ti Derek Liu$^{3}$~~~Evangelos Kalogerakis$^{1,4}$
	\vspace{0.2cm}\\
	$^1$UMass Amherst~~~$^2$École de technologie supérieure~~~$^3$Roblox~~~$^4$TU Crete
	\vspace{-0.2cm}\\
}

\begin{document}

\twocolumn[{%
\renewcommand\twocolumn[1][]{#1}%
\maketitle
\begin{center}
    \centering
    \captionsetup{type=figure}
    \includegraphics[width=\textwidth]{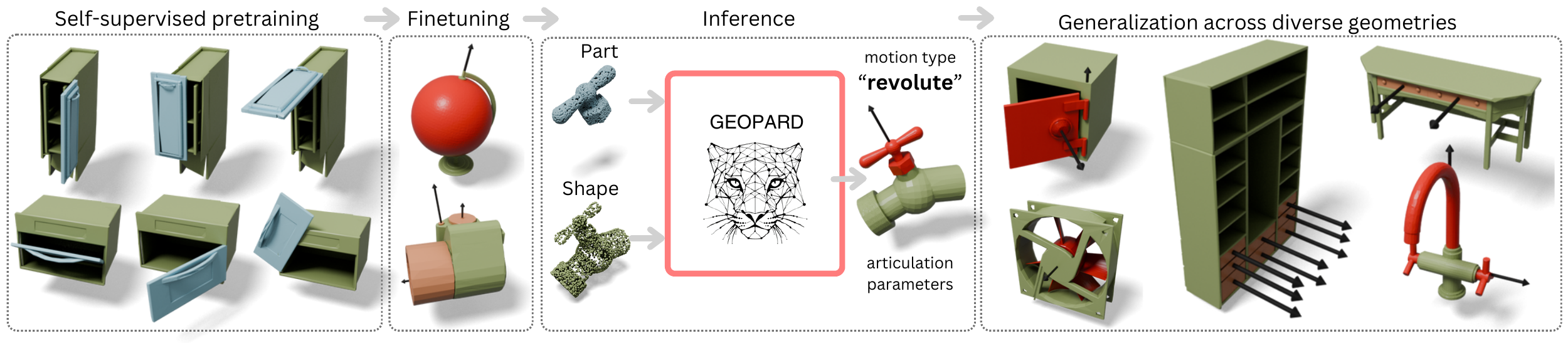}
    \captionof{figure}{\method~allows to predict articulation parameters for diverse object categories and complex kinematic hierarchies. Our method uses \emph{geometrically valid} articulations as a form of self-supervision to pretrain our model, followed by fine-tuning on articulated shape datasets with ground truth annotations, resulting in precise articulation inference. }
    \label{fig:teaser}
\end{center}%
}]

\maketitle
\begin{abstract}
We present \method \footnote{Project page: \href{https://prady1272.github.io/GEOPARD/}{https://prady1272.github.io/GEOPARD/}}, a transformer-based architecture for predicting articulation from a single static snapshot of a 3D shape. The key idea of our method is a pretraining strategy that allows our transformer to learn plausible candidate articulations for 3D shapes based on a geometric-driven search
without manual articulation annotation.  The search automatically discovers physically valid part motions that do not cause detachments or collisions with other shape parts. Our experiments indicate that this geometric pretraining strategy, along with carefully designed choices in our transformer architecture, yields state-of-the-art results in articulation inference in the PartNet-Mobility dataset.
\end{abstract}    
\section{Introduction}
\label{sec:intro}




Articulated objects are pervasive in the physical world—from pieces of furniture and mechanical assemblies to tools and robotic systems. Understanding articulation is critical for building interactive virtual environments and “digital twins” of real‑world objects. As a result, substantial research has been stimulated in articulation understanding and generation for 3D objects \cite{liu2024survey}.


However, a major bottleneck hindering progress is the 
scarcity of training data with detailed articulation annotations, such as motion types, motion axes, centers of rotation for revolute motions, and so on. For example, the most popular benchmark, PartNet-Mobility \cite{xiang2020sapiensimulatedpartbasedinteractive}, includes only a few thousand shapes annotated with articulation parameters.
This restricts the generalizability of existing methods to novel object categories and kinematic hierarchies. Another challenge in articulation understanding is to develop architectures able to accommodate diverse kinematic hierarchies and object categories.

We present \method, a transformer architecture that is able to predict articulation parameters for input shape parts in a diverse set of object categories and complex kinematic hierarchies. Our architecture incorporates the idea of learnable queries, inspired by set transformers \cite{lee2019setr} and modern object detection architectures \cite{carion2020detr} to learn compact, articulation-aware representations for parts. 

In addition, a key idea of our method is to incorporate a pre-training strategy without manual articulation supervision for label-efficient training of our model. Our pretraining strategy
 computes a set of \emph{candidate} articulations based on geometric criteria leading to physically valid articulations without causing
 collision or detachment of parts from the rest of the shape. Using this geometric form of self-supervision,  we pretrain our model, followed by fine-tuning on articulated shape dataset with ground truth annotations to enable precise articulation inference.
Such pretraining is not applicable to several existing architectures (e.g. NAP \cite{NAP}) which necessitate access to the kinematic hierarchy. Our transformer architecture enables effective pretraining on shapes where the kinematic hierarchy is not fixed or known a priori. 
Our method leads to state-of-the-art performance in articulation prediction, as demonstrated by our experiments in the PartNet-Mobility dataset \cite{xiang2020sapiensimulatedpartbasedinteractive}.

In summary, our contributions include:
\begin{itemize}
    \item A pretraining strategy on shape datasets without manual articulation annotations based on a geometric search that automatically discovers candidate articulations for self-supervision.    
    \item A transformer-based network to predict part articulations of shapes based on compact part representations extracted through learnable queries for articulation inference.     
\end{itemize}

\section{Related Work}
\label{sec:related_work}

In this section, we briefly discuss prior methods for articulation inference from various input shape representations.

\paragraph{Multi-State Observation Approaches.} \
Early methods for articulation inference used multiple object states or views to understand part movement. Dearden and Demiris~\cite{dearden2005learning} modeled a 1-DOF robot via a Bayesian network on optical flow features, while Katz~\etal~\cite{katz2008learning} learned planar kinematic models by clustering image features into a kinematic graph. Sturm~\etal~\cite{sturm2008adaptive, sturm2008unsupervised, sturm09rss-manip} proposed probabilistic kinematic models and hierarchies from multi-state inputs. Later work introduced neural networks, such as Deep Part Induction~\cite{yi2018deep} that discovered rigid motions from point cloud pairs. Ditto~\cite{jiang2022ditto} also inferred motion parameters from a pair of object configurations. ScrewNet\cite{jain2021screwnet} applies screw theory to multi-frame depth data, and Liu~\etal~\cite{liu2023building} learn articulable models from 4D sequences by minimizing a motion-based energy function. Qian~\etal~\cite{qian2022understanding} address planar articulation in videos, while Reacto~\cite{song2024reacto} leverages NeRF \cite{mildenhall2020nerf} for in-the-wild videos, learning quasi-sparse skinning weights. Although robust, these methods require sequential or multi-view data, making them impractical for articulation inference in single-snapshot scenarios.

\paragraph{Single-State Observation Approaches.} \noindent Inferring articulation from a single static observation is challenging without motion cues. In 3D, Shape2Motion~\cite{wang2019shape2motion} first jointly estimated part segmentation and motion from a single shape, and RPM-Net~\cite{zihao2019rpmnet} used recurrent networks for modeling displacement sequences from static inputs. Fu~\etal~\cite{fu2024capt} proposed a transformer to regress joint parameters per category from a single point cloud. Image-based methods relied on stronger priors: Li~\etal~\cite{li2020category} predicted part poses and joints from a category-level canonical space, and Jiang~\etal~\cite{jiang2022opd} detect openable parts from an RGB photo. Abdul~\etal~\cite{abdul2022learning} employ a graph-neural network for part segmentation and kinematic hierarchies. Such single-state methods often depend on explicit labels (e.g., part or category labels) to compensate for lack of motion data. In contrast, our method does not enforce part or shape category labels, employs geometric self-supervision to compensate for the scarcity of  annotation data, and handles diverse object categories.

\vspace{-3mm}
\paragraph{Physical Reasoning for Articulation.} \noindent Ensuring physically plausible motions is vital, as part collisions and deformations may break functionality. Weng~\etal~\cite{weng2024twins} optimize collision-free articulations across multi-view RGB-D inputs, while others~\cite{liu2023} penalize self-collisions in hierarchical meshes. Kinematic cues also aid quality: Li~\etal~\cite{li2025dragapart} show a single “drag” point that can reveal part kinematics in images, and MeshArt~\cite{gao2024meshart} generates functional meshes from high-level articulation descriptors. For non-articulated shapes, stability constraints improve generative modeling~\cite{mezghanni2021physicsshape}. Our method incorporates physical reasoning for articulation in the different context of self-supervision for pretraining articulation inference models in 3D shapes.

\vspace{-3mm}
\paragraph{Label-Efficient Articulation Estimation.} \noindent Dense labels for movable parts or motion parameters are costly, prompting label-efficient solutions. GAPartNet~\cite{geng2023gapartnet} uses a small set of actionable parts that generalize across categories, and ScrewNet~\cite{jain2021screwnet} exploits universal 1-DoF joints for category-agnostic articulation. CARTO~\cite{heppert2023carto} learns a joint-agnostic model via physically grounded regularization, and Hartanto~\etal~\cite{hartanto2020hand} employ demonstrations to identify moving parts. Self-supervision can be derived from multi-state observations~\cite{liu2023building, jiang2022opd, sun2024opdmulti, liu2023paris}. Diffusion-based models like CAGE~\cite{liu2024cage} or SINGAPO~\cite{liu2024singapo} learn shape abstractions rather than direct articulation. Liu~\etal~\cite{liu2023semi} rely on semi-weakly supervised learning for articulation inference, and Xu~\etal~\cite{xu2022unsupervised} use semantic category closure. In contrast, our method relies on geometric priors related to physical validity for articulation  for more label-efficient training.

\section{Method}
\label{sec:Method}
\begin{figure*}[th!]
    \centering
    \includegraphics[width=1.0\textwidth]{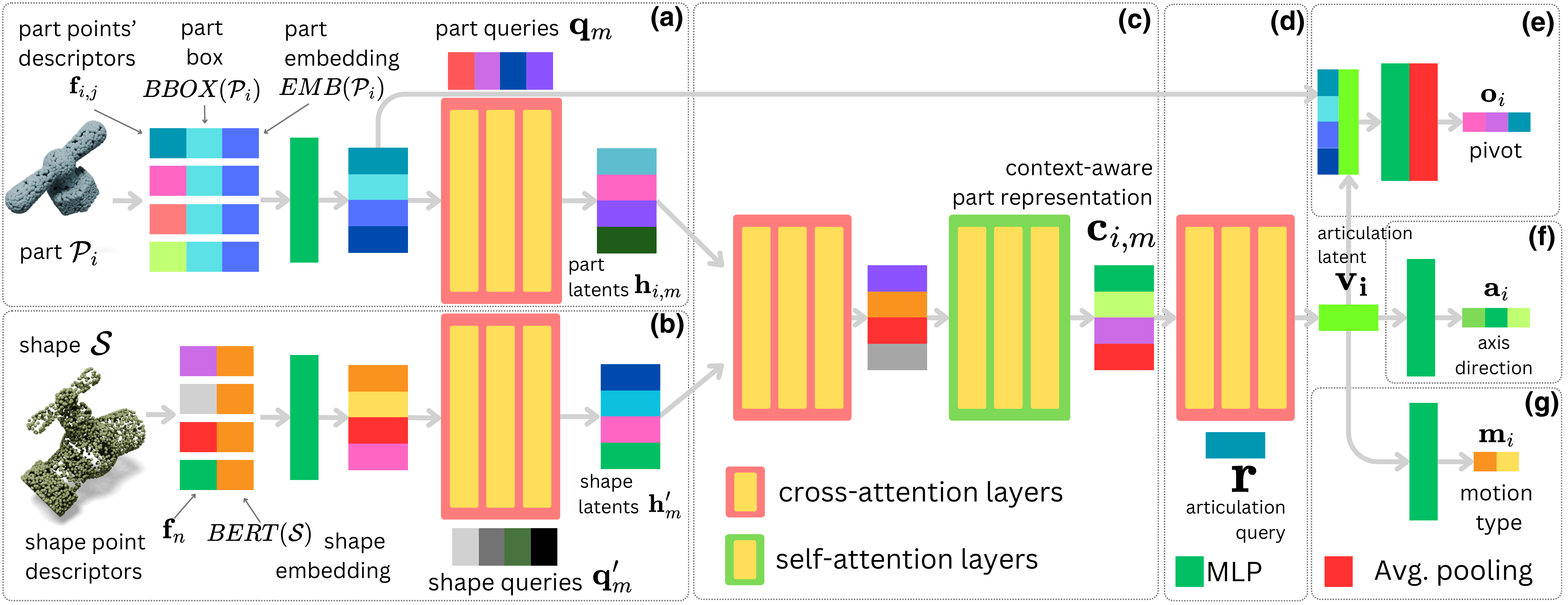}
    \vspace{-3mm}
    \caption{\textbf{\method~overview.} First, we learn 
part feature representations \textbf{(a)} from the part points along with the shape context representation \textbf{(b)}. Second, we enhance the part-level feature representations with the shape context \textbf{(c)}. Third, the representations are aggregated to a compact, articulation-aware part feature vector \textbf{(d)}, which is used to predict the part articulation through a set of three dedicated decoding branches: part pivot prediction \textbf{(e)}; part motion axis prediction \textbf{(f)}; motion type prediction \textbf{(g)}. } 
    \label{fig:architecture} 
   
\end{figure*}

\paragraph{Overview.} 
The goal of \method\ is to predict the articulation of a given set of parts from a single, static snapshot of an input 3D shape. For each part, \method\ assesses whether it is fixed (i.e., it does not articulate), or adheres to a revolute, prismatic, or cylindrical motion. In addition, in the case of an assessed motion for a part, our method predicts the corresponding motion axis. If the motion is revolute or cylindrical, \method\ also predicts the pivot point for the revolute motion. A key idea of our method is to learn articulation-aware features for parts through a transformer-based architecture, also taking into account the whole shape as context. 
Another key idea is a geometric pretraining strategy, where we train the network to predict possible candidate motions of parts extracted through a geometric-based search without any articulation annotations. In the next paragraphs, we first discuss our input assumptions and articulation representation, then we describe our transformer architecture (Section \ref{sec:architecture}), then geometric pretraining along with fine-tuning (Section \ref{sec:pretraining}).

\vspace{-3mm}
\paragraph{Input Assumptions.}  We assume that the object is represented as a point cloud, presegmented into a set of parts, which will be processed by \method\ for assessing their articulation. We also assume that shapes are consistently upright-oriented. We do not impose any restriction on the number of parts or rely on prescribed target kinematic graphs as in prior works \cite{liu2024singapo,liu2024cage,liu2023semiweaklysupervisedobjectkinematic,fu2024capt}. The segmentation may be provided by an external, black-box segmentation model that decomposes the shape into candidate geometric parts (unlabeled), or segments it into parts with semantic labels—our experiments test for both conditions of either labeled or unlabeled input segmentation.

\vspace{-3mm}
\paragraph{Input Representation.} Formally, the input to our method at test time is a shape
$\mS = \{\bp_n\}_{n=1}^{N_s}$, 
where $\bp_n$ is a 3D point position
and $N_s$ is the number of points in the shape. Shapes are normalized according to their longest bounding box axis length, set to $1$, and their centroid is at the coordinate origin.
In addition, we assume all input shapes have consistent upright orientation.  
The shape is composed of a set of parts 
$\{\mP_i\}_{i=1}^{P_s}$, where $P_s$ is the number of parts in the shape. 
Each part is represented as its own set of 3D points $\mP_i=\{\bp_{i,j}\}_{j=1}^{N_{i,s}}$, which is a subset of the original 3D shape,
and $N_{i,s}$ is the number of points in this part.

\paragraph{Articulation Representation.}
The outputs of our method are the predicted articulation parameters of each part $\mP_i$, listed below:
\begin{itemize}
\item \textbf{Motion Type}: 
$\bm_i = [m_{i,rev},\, m_{i,pri}]$, where $m_{i,rev}$ and $m_{i,pri}$ are two binary variables indicating the presence (1) or absence (0) of revolute and prismatic motions respectively. Note that our model can predict that a part adheres to both revolute and prismatic motion, i.e., cylindrical motion (both bits are $1$), or that it is a fixed, non-movable part (both bits are $0$).
\item \textbf{Axis Direction}: $\ba_i \in \mathbb{S}^2$  representing the axis of prismatic, revolute, or cylindrical motion (a unit 3D vector).
\item \textbf{Pivot}: $\bo_i \in \mathbb{R}^3$ specifying the center of rotation, or pivot point, which is used only in the cases of revolute or cylindrical motion. 
\end{itemize}

\subsection{Network Architecture}
\label{sec:architecture}
Our architecture infers articulation parameters for each input part based on (a) the geometric representation of the part and (b) its context --- in our setting, context means where and how the part is placed with respect to the rest of the shape. To this end, we start by learning a 
\emph{part feature representation} (Figure \ref{fig:architecture}a) from the input part points, as well as a \emph{shape context representation} (Figure \ref{fig:architecture}b) from all the shape points. Then we enhance the part-level feature representation with the shape context, leading to a \emph{context-aware part representation} (Figure \ref{fig:architecture}c). 
This representation is further transformed to a compact \emph{part articulation representation}  (Figure \ref{fig:architecture}d) inspired by the learnable query approach used in object detection methods such as DETR \cite{carion2020detr} and set transformers \cite{lee2019setr} -- in our case, the learned query can be thought of as a query to extract the articulation-specific content from the context-aware part representation. 
The resulting part articulation representation is used to predict the part pivot through a dedicated decoding branch (Figure \ref{fig:architecture}e), the part motion axis through another branch  (Figure \ref{fig:architecture}f), and finally the part motion type through another dedicated branch (Figure \ref{fig:architecture}g). We discuss all the above representations and branches  at test time in the following paragraphs.

\paragraph{Part feature representation set.} To build an initial representation for a part, we first encode its point positions. Specifically, for each point $\bp_{i,j}$ belonging to the part $\mP_i$, we pass it through a Fourier-based frequency encoding, commonly used in 3D and other positional encodings \cite{mildenhall2020nerf,zhang20233dshape2vecset}. We concatenate the frequency encoding with the original 3D coordinates to obtain a local $39$-dimensional point descriptor 
\mbox{ $\bff_{i,j} = [PE(\bp_{i,j}), \bp_{i,j}]$ }, where $[\cdot, \cdot]$ denotes concatenation, and $PE$ is a sine and cosine transformation of point coordinates with $12$ different frequencies. 

These point-level descriptors are exclusively local ones, thus we wish to also capture more ``global'', or part-level, information. To this end, we further enhance each of the above point descriptors with: (a) the part-oriented bounding-box representation $BBOX(\mP_i)$ i.e., the 3D part's center location and $3$ box side lengths revealing where it is located in the shape, and how large it is in all three dimensions, and (b) the  $768$-dimensional part embedding $EMB(\mP_i)$ acquired by processing the part points through the PointBERT architecture \cite{yu2022point}, and using its class token i.e., an attention-based aggregation of the part points into a single part representation. Both the bounding box and PointBERT representations helped in our experiments. The scale of parts and position in the shape are correlated with the underlying articulation e.g., cabinet handles are usually elongated in one dimension, and are located on the periphery of the shape. The PointBERT representations are trained in a self-supervised manner for reconstruction of local patches in a large shape dataset (ShapeNet \cite{chang2015shapenet}), and capture structural information useful for part recognition or segmentation. 

If a semantic label is available for the part, we replace the PointBERT representation with a learnable embedding vector representing its part label in the dataset. We discuss the effect of using semantic part label versus PointBERT embeddings in our experiments -- not surprisingly, having access to the label for a part e.g., ``cabinet handle'' helps to recognize its motion type at the expense of an additional, stronger input assumption, which is still commonly used in previous methods. 
After concatenation with the above bounding box and PointBERT or (optionally) semantic label representations, the resulting representations are passed through a shared MLP:
\begin{equation}
\bg_{i,j} = MLP\big( [\bff_{i,j}, BBOX(\mP_i), EMB(\mP_i)] \big)
\end{equation}
where $\bg_{i,j}$ is the representation of the point with index $j$ for the part $\mP_i$. The representation is $D$-dimensional with $D=512$ in our implementation.

One issue with the above point-based representation of the part is that it is sensitive to permutation of its points and is also high-dimensional: given e.g., $N_{i,s}=4096$ points
sampled from our part, the above set representation for it 
 is $4096 \times 512$. Thus, we seek to aggregate it into a more compact part code. Inspired by the learnable query approach of DETR \cite{carion2020detr}, we define a learnable query set of vectors 
$\{\bq_m\}_{m=1}^M$, where $\bq_m$ is a learnable $D$-dimensional query vector ($M=256$ queries in our implementation). Note that the query set is shared across all parts. We use this set of learnable queries in the attention of \cite{vaswani2017attention} to obtain the following feature representation set for the part $\mP_i$:
\begin{align}
\{\bh_{i,m}\}_{m=1}^M = CrossAttn\big( \{\bq_m\}_{m=1}^M, \{ \bg_{i,j} \}_{j=1}^{N_{i,s}} \big)
\end{align}
where the $CrossAttn(\cdot, \cdot)$ is the query-key-value attention of \cite{vaswani2017attention}, or cross-attention, since it operates here on two different sets -- the first argument is the learnable part query set, and the second argument is the point-based set representation of the part. Note that the resulting part feature representation set
$\{\bh_{i,m}\}_{m=1}^M$
is permutation invariant with respect to the part points since it involves an order-insensitive summation over them. 

\vspace{-3mm}
\paragraph{Shape feature set representation.} Our shape feature representations are extracted in a similar manner as the parts. Given each shape point $\bp_n$, we extract a local descriptor \mbox{ $\bff_{n} = [PE(\bp_{n}), \bp_{n}]$ } with the same positional encoding function as above, then enhance it with the PointBERT token global representation $EMB(\mS)$ of the whole shape, and pass it through another MLP to obtain a $D$-dimensional point representation ($D=512$):
\begin{equation}
\bg_{n} = MLP\big( [\bff_{n}, EMB(\mS)] \big)
\end{equation}

Note that we do not use the bounding box representation for shapes, since the shapes are already centered at the origin, and normalized in terms of their scale. We finally compress the point-based set representation of the shape and make it invariant to point permutations through attention on a learnable shape query set $\{\bq'_m\}_{m=1}^M$, shared across all shapes:
\begin{equation}
\{\bh'_{m}\}_{m=1}^M = CrossAttn\big( \{\bq'_m\}_{m=1}^M, \{ \bg_{n} \}_{n=1}^{N_s} \big)
\end{equation}
We note that the learnable shape query set size $M=256$ is the same as the part query set i.e., the shape and part feature representation sets have the same size and dimensionality.

\paragraph{Context-aware part representation.} One issue with our 
part representations so far is that they are completely unaware of the rest of the shape i.e., they are agnostic to how the part is connected, or related to the rest of the shape -- in other words, the context of the part. We use a block of cross-attention layers, where the part feature representations serve as queries and the shape feature representations serve as keys, to obtain contextualized part representations. Then we further process the result through self-attention layers.
\begin{align}
 & \{\bc^{(1)}_{i,m}\}_{m=1}^M = CrossAttn\big( \{\bh_{i,m}\}_{m=1}^M, \{\bh'_{m}\}_{m=1}^M \big)  \\
& \{\bc_{i,m}\}_{m=1}^M = SelfAttn\big( \{\bc^{(1)}_{i,m}\}_{m=1}^M \big) 
\end{align}
where $SelfAttn(\cdot)$ means self-attention and $\{\bc_{i,m}\}_{m=1}^M$ is the resulting context-aware representation set of $M$ vectors for the part $\mP_i$.

\vspace{-2mm}
\paragraph{Part articulation representation.} For predicting the articulation parameters, in our experiments we found that it is more effective to collapse the above part representation set into a single, compact vector representation. This is also useful to reduce the number of parameters for the decoder branches, since the decoder operates on a single vector instead of a set. To perform this collapsing, we follow again the approach of DETR, where we use a single vector $\br$ as a learnable query for articulation (shared across all parts) and the context-aware part feature set $\{\bc_{i,m}\}_{m=1}^M$ as keys in the following cross-attention scheme:
\begin{equation}
\bv_i = CrossAttn\big( \br, \{\bc_{i,m}\}_{m=1}^M \big)
\end{equation}
where $\bv_i$ is the final $D$-dimensional vector representation of the part $\mP_i$ that we use to predict its articulation parameters.

\vspace{-2mm}
\paragraph{Pivot decoder.} In the case of revolute or cylindrical motion for a part, we decode the part articulation vector to the pivot point or center of the rotation. Here, we follow a voting strategy inspired by \cite{fu2024capt}: each point of the part casts a vote for a pivot point. We found this strategy to be much more effective than regressing directly the part articulation vector to the origin. Specifically, we concatenate each of the part's point descriptors $\bg_{i,j}$ with the part articulation representation. Then we regress the resulting point-based representation towards an origin vote (a 3D point) using a fully-connected (FC) block. Finally, we take an average of the origin votes as the final pivot:
\begin{equation}
\bo_i = avg_i \big( FC( [\bg_{i,j}, \bv_i] ) \big)
\end{equation}

\paragraph{Motion Axis decoder.} If the above classification yields one of the two motion types for a part, the motion axis decoder regresses the part articulation representation through another FC block, followed by normalization to ensure that the resulting axis prediction is a unit length vector:
\begin{equation}
\ba_i = Norm \big( FC( \bv_i) \big)
\end{equation}
where $Norm(\cdot)$ is a function dividing the 3D vector output of the FC block by its length.

\paragraph{Motion type decoder.} To decode the part articulation representation to motion type, we use a FC block followed by a sigmoid transformation to predict probabilities for each of the binary random variables $[m_{i,rev},\, m_{i,pri}]$ involved in our motion type representation:
\begin{equation}
[m_{i,rev},\, m_{i,pri}] = \sigma\big( FC( \bv_i) \big)
\end{equation}
where $\sigma(\cdot)$ is a sigmoid function.

\subsection{Geometric Pretraining}
\label{sec:pretraining}
\begin{figure}
    \begin{center}
    \includegraphics[width=1\linewidth]{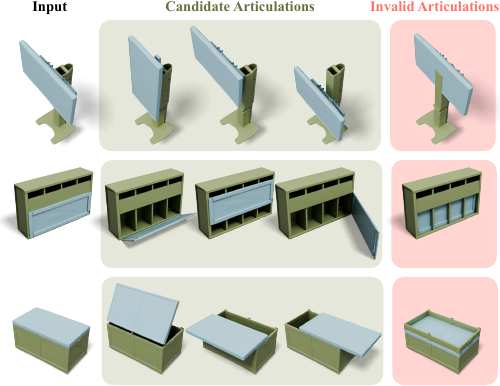}
    \end{center}
    \vspace{-5mm}
    \caption{For a segmented input (left), we compute a set of possible articulations, reject the ones that introduce detachments or collisions with the rest of the part (right), and keep the valid candidate articulations (middle) for our pretraining. }
    \label{fig:PretrainingDataExamples}
\end{figure}
We now discuss pretraining of our architecture on a dataset of shapes segmented into parts \emph{without any articulation annotations}.  Given such a dataset,
the goal of our pretraining is to extract a set of \emph{candidate} articulations for the parts of the training shapes
 based on purely geometric criteria as shown in Figure \ref{fig:PretrainingDataExamples}. 
 We stress that these candidate articulations, which we pretrain our architecture on, do not always coincide with the actual articulations. Yet, it is possible to retrieve a large number of actual articulations from these candidates, making it useful to bootstrap our architecture with the candidate geometric signal we describe in the next paragraphs.
As discussed in our experiments, pretraining on a dataset of segmented shapes, followed by supervised fine-tuning on an articulation-labeled dataset,  significantly improves the performance of our model.

\vspace{-2mm}
\paragraph{Geometric criteria.} If a part is able to revolve about an axis (i.e., revolute motion) or slide along an axis (i.e., prismatic motion) without causing (a) \emph{detachment} of the part from the rest of the shape or (b) \emph{collision} with another part, we consider such articulation as \emph{candidate} for this part. The number of candidate axes and pivot points, however, could still be very large, given that arbitrary motion axes as well as tiny rotations or translations can still 
satisfy the above criteria. Thus, we \emph{prune} many candidate articulations to find the most likely candidate articulations to pretrain our architecture on. 

\vspace{-3mm}
\paragraph{Candidate Axes.}  
Instead of considering all possible axes in $\mathbb{S}^2$ for rotation or translation, we observe that motion axes often coincide with the principal axes of parts. For instance, on average in the training split of Part-Mobility, the articulation axis deviates from one of the PCA axes by only $3.5\%$. Thus, we perform PCA on densely sampled points of the parts of our pretraining dataset to extract candidate axes for either translation or rotation. 

\vspace{-3mm}
\paragraph{Candidate Pivots.}  
For revolute motion, candidate pivot points are also required. To this end, we compute the axis-aligned bounding box of each part in our pretraining dataset and extract its 8 vertices along with the part centroid.


\vspace{-3mm}
\paragraph{Motion Range Pruning.}  
For \emph{prismatic motion}, each candidate axis defines a translation direction. Translations are permitted up to the maximum extent $L$ of the largest axis of the oriented bounding box of the part. We discard any prismatic articulation with a very small range less than $\epsilon = L/10$. For \emph{revolute motion}, we form candidate pairs by combining each candidate axis (3 possible axes from PCA) with every candidate pivot ($9$ pivots). This yields up to $3 \times 9$ candidate articulations for revolute motion. To ensure sufficient rotation range, we remove any candidate that allows a rotation limit below $\omega=90^\circ$.

\vspace{-3mm}
\paragraph{Collision and Detachment Pruning.}  
We simulate the translation and rotation defined by each candidate articulation for each part up to the limits described above. 
We employ the Expanding Polytope Algorithm (EPA) \cite{van2001proximity} on the mesh triangles to detect potential collisions, allowing a tolerance threshold $\epsilon'=0.01$ for minor contacts. The same threshold is also applied to allow unintended detachments.

\vspace{-3mm}
\paragraph{Pivot pruning.}  
Since multiple candidate pivot points may still exist after the above pruning criteria, we select the rotation pivot that yields the largest rotation range per axis.

\vspace{-3mm}
\paragraph{Pretraining adjustments.}  
The result of the above geometry-based search is a set of likely candidate articulations per part in the pretraining dataset. This can include either a prismatic articulation, a revolute articulation, both, or none (if pruning rejects all initial candidate articulations). We note that a part might be associated with more than one rotation axis i.e., up to 3 axes along with their associated pivots. To this end, we expand the decoder MLPs to predict articulation parameters along the three axes. We order the axes based on their corresponding eigenvalue from PCA. We observed that ordering by eigenvalue instead of rotation range or prismatic range significantly improved training. 

\vspace{-3mm}
\paragraph{Losses.}  
Given the final candidate articulation parameters extracted through the above geometric criteria, we pretrain our architecture such that the discrepancy between motion axes, pivot points, and motion types is minimized. Specifically, we use binary cross entropy $L_{ce}$ for each of two motion types, cosine similarity $L_{cos}$ for the motion axes (masked for parts deemed as fixed), and $L_1$ loss for pivot points (masked for parts with no revolute motion). The combined loss for training is the weighted sum of these individual losses: \mbox{$L=L_{ce}+\lambda_1 \cdot L_{cos}+\lambda_2 \cdot L_1$},
where $\lambda_1=1$, $\lambda_2=1$ in our implementation. 

\vspace{-3mm}
\paragraph{Supervised fine-tuning.} After pretraining, we fine-tune our model using articulation supervision (Part-Mobility \cite{xiang2020sapien} in our experiments). 
We note that Part-Mobility allows one degree of freedom (one‑DoF) axis, thus, we use the original base architecture, reinitializing the decoders. We restart the training, with the rest of the model, using the same loss as in pretraining.

\vspace{-3mm}
\paragraph{Implementation details.} 
 The model is trained using the AdamW optimizer~\cite{loshchilov2017decoupled}  for $1000$ epochs (out of which $500$ are used for pretraining). Our pretraining uses a learning rate of $10^{-4}$. For fine-tuning, we use a learning rate of $10^{-5}$.

\section{Experiments}
\label{sec:Experiment}
\label{sec:experiments}
\label{sec:results}
\vspace{-2mm}
We now discuss the experimental validation of our method. We first present details about our experimental setup, then we discuss competing methods, results, and our ablation.

\begin{table}[t!]
\centering
\setlength{\tabcolsep}{3pt}
\begin{tabular}{l|ccccc}
\toprule
\textbf{Model} & \textbf{AE $\downarrow$} & \textbf{PE $\downarrow$} & \textbf{R-ACC $\uparrow$} & \textbf{P-ACC $\uparrow$} & \textbf{AD $\downarrow$} \\
\midrule
CAGE  & 11.41& 0.09& \textbf{0.98} & \textbf{0.98}  & 0.07 \\
SINGAPO     &12.15&0.10&0.97&0.97 & 0.07\\
\method       & \textbf{8.87} & \textbf{0.06} & \textbf{0.98} & \textbf{0.98}& \textbf{0.06} \\
\bottomrule
\end{tabular}
\vspace{-3mm}
\caption{Evaluation in the \emph{labeled} part condition}
\label{table:comp_quant_eval_labeled}
\end{table}

\begin{table}[t!]
\centering
\setlength{\tabcolsep}{3pt}
\begin{tabular}{l|ccccc@{}}
\toprule
\textbf{Model} & \textbf{AE $\downarrow$} & \textbf{PE $\downarrow$} & \textbf{R-ACC $\uparrow$} & \textbf{P-ACC $\uparrow$} & \textbf{AD $\downarrow$} \\
\midrule
CAGE-u &  12.83& 0.11& 0.89& 0.89 & 0.13 \\
SINGAPO-u  &11.20&0.12&0.90&0.91 & 0.11\\
\method-u & \textbf{9.18} & \textbf{0.05} & \textbf{0.92}  & \textbf{0.93} & \textbf{0.09} \\
\bottomrule
\end{tabular}
\vspace{-3mm}
\caption{Evaluation in the \emph{unlabeled} part condition}
\label{table:comp_quant_eval_unlabeled}
\vspace{-3mm}
\end{table}




\begin{figure}
    \vspace{-3mm}
    \begin{center}
    \includegraphics[width=1\linewidth]{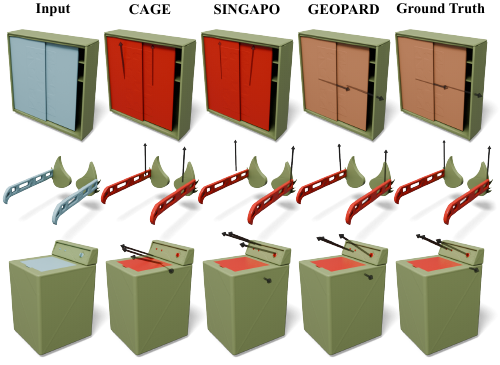}
    \end{center}
    \vspace{-6mm}
   \caption{\textbf{Qualitative Comparisons (with labels)}
\textcolor{revolute}{\rule{6pt}{6pt}} indicates parts predicted or labeled as \textbf{revolute},     
\textcolor{prismatic}{\rule{6pt}{6pt}} indicates parts predicted or labeled as \textbf{prismatic}, 
\textcolor{candidate}{\rule{6pt}{6pt}} denotes \textbf{input parts}. Results showcase that our model predicts motion type and axis direction (\textbf{Row 1}) and revolute points (\textbf{Rows 2-3}) with improved performance.}
    \vspace{-3mm}
    \label{fig:QualitativeLabel}
\end{figure}

\begin{figure*}
    \vspace{-5mm}
    \begin{center}
    \includegraphics[width=1\linewidth]{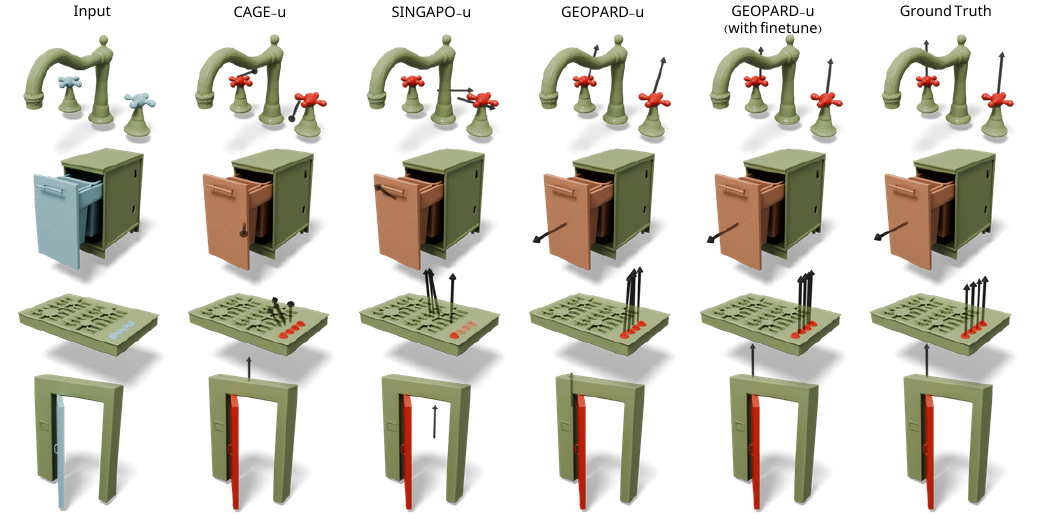}
    \end{center}
    \vspace{-6mm}
    \caption{\textbf{Qualitative comparisons (without labels).}
\textcolor{revolute}{\rule{6pt}{6pt}} are parts predicted or labeled as \textbf{revolute},     
\textcolor{prismatic}{\rule{6pt}{6pt}} are parts predicted or labeled as \textbf{prismatic}, 
\textcolor{candidate}{\rule{6pt}{6pt}} are \textbf{input parts}. Predicted axes are shown with an arrow ($\uparrow$). While baselines based on part abstractions struggle to predict plausible articulation parameters, our base model using fine grained point features, produces articulation parameters closely matching the ground truth - which are further enhanced by our pretraining strategy, supplying geometric and articulation priors refined during fine-tuning.}
    \label{fig:QualitativePointbert}
\end{figure*}

\vspace{-2mm}
\paragraph{Experimental setup.}
We perform experiments on the PartNet-Mobility dataset~\cite{xiang2020sapiensimulatedpartbasedinteractive}, which includes a diverse collection of object categories with annotated articulation parameters. In our experiments, we utilize all 46 categories.  We use the test split provided by ~\cite{liu2023semiweaklysupervisedobjectkinematic}. Specifically, we use $1830$ shapes for training, 
and $274$ shapes for testing. 
We use the same set of shapes as that of the training split for pretraining, yet relying only on segmented parts without using the original articulation annotation. We performed experiments in two conditions: (a) the ``unlabeled'' condition where segmented parts have no available semantic labels (i.e., a geometric decomposition), (b) the ``labeled'' condition where parts are labeled (a stronger assumption in the input data). In the case of label inputs, our method uses their part label embeddings instead of the PointBERT part embeddings, as discussed in Section \ref{sec:architecture}. 
For the rest of this section, the reported evaluation is based on the part segmentations provided in PartNet-Mobility. The supplement includes evaluation when an off‑the‑shelf segmentation model (PartSTAD \cite{kim2024partstad}) is used instead.


\vspace{-4mm}
\paragraph{Competing methods.}
We compare \method~ with two state-of-the-art articulation models, \textbf{CAGE} \cite{liu2024cage} and \textbf{SINGAPO} \cite{liu2024singapo}. Like \method, both methods support cross-category training, do not require specific kinematic graphs as part of their input, and are not constrained to operate within a single category. 
They also assume that input shapes are segmented into labeled parts. 
We note that the original CAGE method supports a maximum of $32$ parts per shape and $7$ categories. We increase this limit to $116$ to accommodate shapes in our dataset as well as the $46$ categories. We also developed two variants of the above methods, called \textbf{CAGE-u} and \textbf{SINGAPO-u}. For both variants, we replace their semantic part label input with the same PointBERT part embeddings used in our method. Similarly, we include results for two variants of our method -- the variant in the unlabeled condition is called \textbf{``\method-u''}. All methods are trained and tested on the same splits.

\vspace{-4mm}
\paragraph{Metrics.}
We evaluate our approach using a set of metrics that assess the orientation and positional accuracy of the predicted articulation parameters as well as their motion type. Specifically, we use: (a)  the \textbf{Axis Error (AE)} measured as the angular deviation (in degrees) between the predicted motion axis and the corresponding ground-truth axis, (b) 
for revolute joints, we further measure the \textbf{Point Error (PE)} as the perpendicular distance from the predicted pivot point to the ground-truth rotation axis, as any point along the ground-truth axis is valid, (c) the \textbf{Revolute Accuracy (R-ACC)} and \textbf{Prismatic Accuracy (P-ACC)}, which denote the percentages of correctly predicted binary labels for revolute and prismatic motions, respectively. Additionally, we report \textbf{Articulated Distance (AD)}—a metric designed to capture the discrepancy between predicted and ground-truth articulations across sampled motion states. Specifically, we compute the average Chamfer Distance between the ground-truth and predicted articulated point clouds of the parts across five sampled motion states.

\vspace{-3mm}
\paragraph{Results.} Table \ref{table:comp_quant_eval_labeled} presents the evaluation metrics for the competing methods in the ``labeled'' condition. Our method outperforms both CAGE and SINGAPO in terms of axis error and pivot point. Relative to CAGE, it improves AE by $22\%$, PE by $33\%$, and AD by $14\%$. Relative to SINGAPO, it improves AE by $27\%$, PE by $40\%$ and AD by $14\%$. All methods have comparable performance in terms of recognizing motion type, which is expected given the availability of input semantic labels.

 Table \ref{table:comp_quant_eval_unlabeled} reports results for the unlabeled condition. Our method outperforms both the CAGE and SINGAPO variants according to all metrics. The relative improvements are similar to the labeled condition in terms of AE and PE, 
 while we also observe improvements in motion type prediction and the Articulated Distance. 
 As expected, the performance for all methods is worse without using labels, especially for motion type recognition. 

Figures \ref{fig:QualitativeLabel} and \ref{fig:QualitativePointbert} showcase results for labeled and unlabeled settings, respectively. When comparing results in the labeled condition, Figure \ref{fig:QualitativeLabel}, we consistently predict articulation parameters that are more plausible, such as motion type and axis (row 1)  and revolute origin (row 3). When comparing results in unlabeled condition, Figure \ref{fig:QualitativePointbert}, our method still generalizes well to complex geometries that require fine-grained representation e.g., example handles of a faucet (row 1); produces consistent results for small parts like oven knobs (row 3); and correctly estimates pivot points where baselines fail (row 4).


\begin{table}[t!]
\vspace{-3mm}
\centering
\begin{tabular}{l@{}|@{}cccc@{}}
\toprule
\textbf{Model} & \textbf{AE $\downarrow$} & \textbf{PE $\downarrow$} & \textbf{R-ACC $\uparrow$} & \textbf{P-ACC $\uparrow$} \\
\midrule
\method-u-nopr    &     10.67& 0.06& 0.87&0.88 \\
\method-u & \textbf{9.18} & \textbf{0.05} & \textbf{0.92} & \textbf{0.93} \\
\bottomrule
\end{tabular}
\vspace{-3mm}
\caption{Ablation study for pretraining (\emph{unlabeled} condition).}
\label{table:ablation_pretraining}
\vspace{-5mm}
\end{table}

\begin{table}[t!]
\centering
\begin{tabular}{lc|cc} 
\toprule
\textbf{Model} & \textbf{Part aggr.} & \textbf{AE $\downarrow$} & \textbf{PE $\downarrow$} \\
\midrule
Point  & Avg. & 11.42& 0.11 \\ 
Point  & CA   & \textbf{10.56} & 0.09 \\ 
Displ. & CA   & 10.67& \textbf{0.06} \\ 
\bottomrule
\end{tabular}
\caption{Ablation on motion parameter prediction strategies (\emph{unlabeled} condition, \emph{no pretraining}).}
\vspace{-6mm}
\label{table:ablation_motion}
\end{table}

\vspace{-3mm}
\paragraph{Ablation.}
Table \ref{table:ablation_pretraining} presents our evaluation metrics for \method\ when pretraining is not used (``\method-u-nopr'' variant) versus when it is used in the unlabeled condition. Evaluation metrics are improved when pretraining is used. We see a $14\%$ relative decrease in AE, $16\%$ relative decrease in PE. We also observe a $5\%$ improvement in motion type recognition. 

In terms of architectural choices, we also study the effectiveness of using a dedicated articulation learnable query (Table \ref{table:ablation_motion}, row 2)  for aggregating context-aware part features
versus mean-pooling 
(Table \ref{table:ablation_motion}, row 1). We see a $7.5\%$ relative decrease in AE, $18\%$ relative decrease in PE. We then demonstrate the benefit of regressing per-point displacements (Table \ref{table:ablation_motion}, row 3) compared to 
directly predicting the origin point from the articulation token (Table \ref{table:ablation_motion}, row 2). We see a minor $1\%$ relative increase in AE, yet a significant further $33\%$ relative decrease in PE.
Our results suggest that a combination of displacement regression for pivot prediction and part latent aggregation via cross-attention (CA) improves prediction results, especially in terms of point error. We note that the motion type accuracies remain comparable for the above architectural choices.
\section{Conclusion and Limitations}
\vspace{-2mm}
We discussed a model for articulation prediction in 3D shapes based on compact part representations extracted through learnable queries in neural attention, as well as a geometric pretraining strategy aiming to discover physically valid articulations free of collisions or detachments. 

\begin{wrapfigure}{R}{1.4in}
    \vspace{-10pt}
    \begin{minipage}[b]{1\linewidth}
        \includegraphics[width=1.1\linewidth, trim={5mm 2mm 0mm 0mm}]{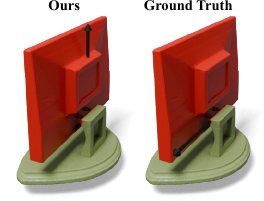}
        \caption{\textbf{Failure case.} Instead of predicting the ground-truth motion axis, our method may sometimes predict one of the candidate axes used for pretraining.}
        \vspace{-4mm}
        \label{fig:failure_case}
    \end{minipage}
\end{wrapfigure}
Despite improving the state-of-the-art in articulation prediction, there are still several limitations. First, the geometric pretraining does not guarantee the discovery of all correct part articulations. The validity of articulations is checked only through detachment and collision, while other approaches, e.g, VLMs could also provide useful feedback. After pre-training, fine-tuning may result in predicting the candidate articulations instead of the ground-truth ones (see the inset Figure \ref{fig:failure_case}). Finally, our method still relies on the availability of unlabeled or labeled part segmentations. Learning to infer moving part segmentations in a zero-shot or self-supervised manner remains an open problem.

\vspace{-5mm}
\paragraph{Acknowledgements.}
This project has received funding from the European Research Council (ERC) under the Horizon research and innovation programme (Grant agreement No. 101124742).

{
    \small
    \bibliographystyle{ieeenat_fullname}
    \bibliography{main}

@String(CVPR= {IEEE Conf. Comput. Vis. Pattern Recog.})

@String(ICCV= {Int. Conf. Comput. Vis.})

@String(ECCV= {Eur. Conf. Comput. Vis.})

@String(TOG= {ACM Trans. Graph.})

@String(IJCAI = {IJCAI})

@String(CVPR  = {CVPR})

@String(ICCV  = {ICCV})

@String(ECCV  = {ECCV})

@String(TOG   = {ACM TOG})

@article{zhang20233dshape2vecset,
  title={3dshape2vecset: A 3d shape representation for neural fields and generative diffusion models},
  author={Zhang, Biao and Tang, Jiapeng and Niessner, Matthias and Wonka, Peter},
  journal={ACM Transactions on Graphics (TOG)},
  volume={42},
  number={4},
  pages={1--16},
  year={2023},
  publisher={ACM New York, NY, USA}
}

@misc{xiang2020sapiensimulatedpartbasedinteractive,
      title={SAPIEN: A SimulAted Part-based Interactive ENvironment}, 
      author={Fanbo Xiang and Yuzhe Qin and Kaichun Mo and Yikuan Xia and Hao Zhu and Fangchen Liu and Minghua Liu and Hanxiao Jiang and Yifu Yuan and He Wang and Li Yi and Angel X. Chang and Leonidas J. Guibas and Hao Su},
      year={2020},
      eprint={2003.08515},
      archivePrefix={arXiv},
      primaryClass={cs.CV},
      url={https://arxiv.org/abs/2003.08515}, 
}

@misc{liu2023semiweaklysupervisedobjectkinematic,
      title={Semi-Weakly Supervised Object Kinematic Motion Prediction}, 
      author={Gengxin Liu and Qian Sun and Haibin Huang and Chongyang Ma and Yulan Guo and Li Yi and Hui Huang and Ruizhen Hu},
      year={2023},
      eprint={2303.17774},
      archivePrefix={arXiv},
      primaryClass={cs.CV},
      url={https://arxiv.org/abs/2303.17774}, 
}

@article{liu2024survey,
  title={Survey on Modeling of Human-made Articulated Objects},
  author={Liu, Jiayi and Savva, Manolis and Mahdavi-Amiri, Ali},
  journal={arXiv preprint arXiv:2403.14937},
  year={2024}
}

@article{NAP,
  author       = {Jiahui Lei and
                  Congyue Deng and
                  Bokui Shen and
                  Leonidas J. Guibas and
                  Kostas Daniilidis},
  title        = {{NAP:} Neural 3D Articulation Prior},
  journal      = {CoRR},
  volume       = {abs/2305.16315},
  year         = {2023},
  url          = {https://doi.org/10.48550/arXiv.2305.16315},
  doi          = {10.48550/ARXIV.2305.16315},
  eprinttype    = {arXiv},
  eprint       = {2305.16315},
}

@inproceedings{van2001proximity,
  title={Proximity queries and penetration depth computation on 3d game objects},
  author={Van Den Bergen, Gino},
  booktitle={Game developers conference},
  volume={170},
  pages={209},
  year={2001}
}

@InProceedings{xiang2020sapien,
author = {Xiang, Fanbo and Qin, Yuzhe and Mo, Kaichun and Xia, Yikuan and Zhu, Hao and Liu, Fangchen and Liu, Minghua and Jiang, Hanxiao and Yuan, Yifu and Wang, He and Yi, Li and Chang, Angel X. and Guibas, Leonidas J. and Su, Hao},
title = {{SAPIEN}: A SimulAted Part-based Interactive ENvironment},
booktitle = {The IEEE Conference on Computer Vision and Pattern Recognition (CVPR)},
month = {June},
year = {2020}}

@inproceedings{carion2020detr,
  title={End-to-end object detection with transformers},
  author={Carion, Nicolas and Massa, Francisco and Synnaeve, Gabriel and Usunier, Nicolas and Kirillov, Alexander and Zagoruyko, Sergey},
  booktitle={European conference on computer vision},
  pages={213--229},
  year={2020},
  organization={Springer}
}

@inproceedings{lee2019setr,
  title={Set transformer: A framework for attention-based permutation-invariant neural networks},
  author={Lee, Juho and Lee, Yoonho and Kim, Jungtaek and Kosiorek, Adam and Choi, Seungjin and Teh, Yee Whye},
  booktitle={International conference on machine learning},
  pages={3744--3753},
  year={2019},
  organization={PMLR}
}

@inproceedings{mildenhall2020nerf,
  title={NeRF: Representing Scenes as Neural Radiance Fields for View Synthesis},
  author={Mildenhall, Ben and Srinivasan, Pratul P and Tancik, Matthew and Barron, Jonathan T and Ramamoorthi, Ravi and Ng, Ren},
  booktitle={European Conference on Computer Vision},
  pages={405--421},
  year={2020},
  organization={Springer}
}

@inproceedings{heppert2023carto,
  title={Carto: Category and joint agnostic reconstruction of articulated objects},
  author={Heppert, Nick and Irshad, Muhammad Zubair and Zakharov, Sergey and Liu, Katherine and Ambrus, Rares Andrei and Bohg, Jeannette and Valada, Abhinav and Kollar, Thomas},
  booktitle={Proceedings of the IEEE/CVF Conference on Computer Vision and Pattern Recognition},
  pages={21201--21210},
  year={2023}
}

@article{zihao2019rpmnet,
author = {Yan, Zihao and Hu, Ruizhen and Yan, Xingguang and Chen, Luanmin and Van Kaick, Oliver and Zhang, Hao and Huang, Hui},
title = {RPM-Net: recurrent prediction of motion and parts from point cloud},
year = {2019},
journal = {ACM Transactions on Graphics (Proceedings of SIGGRAPH Asia)},
issue_date = {December 2019},
publisher = {Association for Computing Machinery},
address = {New York, NY, USA},
volume = {38},
number = {6},
issn = {0730-0301},
url = {https://doi.org/10.1145/3355089.3356573},
doi = {10.1145/3355089.3356573},
month = nov,
articleno = {240},
numpages = {15}
}

@article{liu2024singapo,
  title={SINGAPO: Single Image Controlled Generation of Articulated Parts in Objects},
  author={Liu, Jiayi and Iliash, Denys and Chang, Angel X and Savva, Manolis and Mahdavi-Amiri, Ali},
  journal={arXiv preprint arXiv:2410.16499},
  year={2024}
}

@InProceedings{ sturm09rss-manip,
	author = {J. Sturm and C. Stachniss and V. Pradeep and C. Plagemann and K. Konolige and W. Burgard},
	title = "Towards Understanding Articulated Objects",
	booktitle = "Proc. of the Workshop on Robot Manipulation at Robotics: Science and Systems Conference (RSS)",
	year = "2009",
	month= "June",
titleurl = {sturm09rss-manip.pdf},
topic={articulated-objects},
}

@inproceedings{dearden2005learning,
  title={Learning forward models for robots},
  author={Dearden, Anthony and Demiris, Yiannis},
  booktitle={IJCAI},
  volume={5},
  pages={1440},
  year={2005}
}

@inproceedings{sturm2008adaptive,
  title={Adaptive body scheme models for robust robotic manipulation.},
  author={Sturm, J{\"u}rgen and Plagemann, Christian and Burgard, Wolfram},
  booktitle={Robotics: Science and systems},
  year={2008},
  organization={Zurich}
}

@inproceedings{sturm2008unsupervised,
  title={Unsupervised body scheme learning through self-perception},
  author={Sturm, Jurgen and Plagemann, Christian and Burgard, Wolfram},
  booktitle={2008 IEEE International Conference on Robotics and Automation},
  pages={3328--3333},
  year={2008},
  organization={IEEE}
}

@inproceedings{katz2008learning,
author = {Katz, Dov and Pyuro, Yuri and Brock, Oliver},
year = {2008},
month = {06},
pages = {},
booktitle = {Proceedings of Robotics: Science and Systems IV},
title = {Learning to Manipulate Articulated Objects in Unstructured Environments Using a Grounded Relational Representation},
journal = {Proceedings of Robotics: Science and Systems IV},
doi = {10.15607/RSS.2008.IV.033}
}

@article{yi2018deep,
author = {Yi, Li and Huang, Haibin and Liu, Difan and Kalogerakis, Evangelos and Su, Hao and Guibas, Leonidas},
title = {Deep part induction from articulated object pairs},
year = {2018},
issue_date = {December 2018},
publisher = {Association for Computing Machinery},
address = {New York, NY, USA},
volume = {37},
number = {6},
issn = {0730-0301},
url = {https://doi.org/10.1145/3272127.3275027},
doi = {10.1145/3272127.3275027},
journal = {ACM Trans. Graph.},
month = dec,
articleno = {209},
numpages = {15}
}

@inproceedings{jiang2022ditto,
  title={Ditto: Building digital twins of articulated objects from interaction},
  author={Jiang, Zhenyu and Hsu, Cheng-Chun and Zhu, Yuke},
  booktitle={Proceedings of the IEEE/CVF Conference on Computer Vision and Pattern Recognition},
  pages={5616--5626},
  year={2022}
}

@inproceedings{jain2021screwnet,
  title={Screwnet: Category-independent articulation model estimation from depth images using screw theory},
  author={Jain, Ajinkya and Lioutikov, Rudolf and Chuck, Caleb and Niekum, Scott},
  booktitle={2021 IEEE International Conference on Robotics and Automation (ICRA)},
  pages={13670--13677},
  year={2021},
  organization={IEEE}
}

@inproceedings{liu2023building,
  title={Building rearticulable models for arbitrary 3d objects from 4d point clouds},
  author={Liu, Shaowei and Gupta, Saurabh and Wang, Shenlong},
  booktitle={Proceedings of the IEEE/CVF Conference on Computer Vision and Pattern Recognition},
  pages={21138--21147},
  year={2023}
}

@inproceedings{qian2022understanding,
  title={Understanding 3d object articulation in internet videos},
  author={Qian, Shengyi and Jin, Linyi and Rockwell, Chris and Chen, Siyi and Fouhey, David F},
  booktitle={Proceedings of the IEEE/CVF Conference on Computer Vision and Pattern Recognition},
  pages={1599--1609},
  year={2022}
}

@inproceedings{song2024reacto,
  title={Reacto: Reconstructing articulated objects from a single video},
  author={Song, Chaoyue and Wei, Jiacheng and Foo, Chuan Sheng and Lin, Guosheng and Liu, Fayao},
  booktitle={Proceedings of the IEEE/CVF Conference on Computer Vision and Pattern Recognition},
  pages={5384--5395},
  year={2024}
}

@inproceedings{wang2019shape2motion,
  title={Shape2motion: Joint analysis of motion parts and attributes from 3d shapes},
  author={Wang, Xiaogang and Zhou, Bin and Shi, Yahao and Chen, Xiaowu and Zhao, Qinping and Xu, Kai},
  booktitle={Proceedings of the IEEE/CVF Conference on Computer Vision and Pattern Recognition},
  pages={8876--8884},
  year={2019}
}

@inproceedings{fu2024capt,
  title={CAPT: Category-level Articulation Estimation from a Single Point Cloud Using Transformer},
  author={Fu, Lian and Ishikawa, Ryoichi and Sato, Yoshihiro and Oishi, Takeshi},
  booktitle={2024 IEEE International Conference on Robotics and Automation (ICRA)},
  pages={751--757},
  year={2024},
  organization={IEEE}
}

@inproceedings{li2020category,
  title={Category-level articulated object pose estimation},
  author={Li, Xiaolong and Wang, He and Yi, Li and Guibas, Leonidas J and Abbott, A Lynn and Song, Shuran},
  booktitle={Proceedings of the IEEE/CVF conference on computer vision and pattern recognition},
  pages={3706--3715},
  year={2020}
}

@inproceedings{jiang2022opd,
  title={OPD: Single-view 3d openable part detection},
  author={Jiang, Hanxiao and Mao, Yongsen and Savva, Manolis and Chang, Angel X},
  booktitle={European Conference on Computer Vision},
  pages={410--426},
  year={2022},
  organization={Springer}
}

@inproceedings{geng2023gapartnet,
  title={Gapartnet: Cross-category domain-generalizable object perception and manipulation via generalizable and actionable parts},
  author={Geng, Haoran and Xu, Helin and Zhao, Chengyang and Xu, Chao and Yi, Li and Huang, Siyuan and Wang, He},
  booktitle={Proceedings of the IEEE/CVF Conference on Computer Vision and Pattern Recognition},
  pages={7081--7091},
  year={2023}
}

@inproceedings{liu2023semi,
  title={Semi-weakly supervised object kinematic motion prediction},
  author={Liu, Gengxin and Sun, Qian and Huang, Haibin and Ma, Chongyang and Guo, Yulan and Yi, Li and Huang, Hui and Hu, Ruizhen},
  booktitle={Proceedings of the IEEE/CVF Conference on Computer Vision and Pattern Recognition},
  pages={21726--21735},
  year={2023}
}

@inproceedings{xu2022unsupervised,
  title={Unsupervised kinematic motion detection for part-segmented 3d shape collections},
  author={Xu, Xianghao and Ruan, Yifan and Sridhar, Srinath and Ritchie, Daniel},
  booktitle={ACM SIGGRAPH 2022 Conference Proceedings},
  pages={1--9},
  year={2022}
}

@inproceedings{liu2024cage,
  title={CAGE: controllable articulation generation},
  author={Liu, Jiayi and Tam, Hou In Ivan and Mahdavi-Amiri, Ali and Savva, Manolis},
  booktitle={Proceedings of the IEEE/CVF Conference on Computer Vision and Pattern Recognition},
  pages={17880--17889},
  year={2024}
}

@inproceedings{hartanto2020hand,
  title={Hand-motion-guided articulation and segmentation estimation},
  author={Hartanto, Richard Sahala and Ishikawa, Ryoichi and Roxas, Menandro and Oishi, Takeshi},
  booktitle={2020 29th IEEE International Conference on Robot and Human Interactive Communication (RO-MAN)},
  pages={807--813},
  year={2020},
  organization={IEEE}
}

@inproceedings{sun2024opdmulti,
  title={OPDMulti: Openable part detection for multiple objects},
  author={Sun, Xiaohao and Jiang, Hanxiao and Savva, Manolis and Chang, Angel},
  booktitle={2024 International Conference on 3D Vision (3DV)},
  pages={169--178},
  year={2024},
  organization={IEEE}
}

@inproceedings{liu2023paris,
  title={Paris: Part-level reconstruction and motion analysis for articulated objects},
  author={Liu, Jiayi and Mahdavi-Amiri, Ali and Savva, Manolis},
  booktitle={Proceedings of the IEEE/CVF International Conference on Computer Vision},
  pages={352--363},
  year={2023}
}

@inproceedings{abdul2022learning,
  title={Learning to infer kinematic hierarchies for novel object instances},
  author={Abdul-Rashid, Hameed and Freeman, Miles and Abbatematteo, Ben and Konidaris, George and Ritchie, Daniel},
  booktitle={2022 International Conference on Robotics and Automation (ICRA)},
  pages={8461--8467},
  year={2022},
  organization={IEEE}
}

@InProceedings{weng2024twins,
    author    = {Weng, Yijia and Wen, Bowen and Tremblay, Jonathan and Blukis, Valts and Fox, Dieter and Guibas, Leonidas and Birchfield, Stan},
    title     = {Neural Implicit Representation for Building Digital Twins of Unknown Articulated Objects},
    booktitle = {Proceedings of the IEEE/CVF Conference on Computer Vision and Pattern Recognition (CVPR)},
    month     = {June},
    year      = {2024},
    pages     = {3141-3150}
}

@InProceedings{liu2023,
    author    = {Liu, Xueyi and Wang, Bin and Wang, He and Yi, Li},
    title     = {Few-Shot Physically-Aware Articulated Mesh Generation via Hierarchical Deformation},
    booktitle = {Proceedings of the IEEE/CVF International Conference on Computer Vision (ICCV)},
    month     = {October},
    year      = {2023},
    pages     = {854-864}
}

@InProceedings{li2025dragapart,
author="Li, Ruining
and Zheng, Chuanxia
and Rupprecht, Christian
and Vedaldi, Andrea",
editor="Leonardis, Ale{\v{s}}
and Ricci, Elisa
and Roth, Stefan
and Russakovsky, Olga
and Sattler, Torsten
and Varol, G{\"u}l",
title="DragAPart: Learning a Part-Level Motion Prior for Articulated Objects",
booktitle="Computer Vision -- ECCV 2024",
year="2025",
publisher="Springer Nature Switzerland",
address="Cham",
pages="165--183",
}

@InProceedings{mezghanni2021physicsshape,
    author    = {Mezghanni, Mariem and Boulkenafed, Malika and Lieutier, Andre and Ovsjanikov, Maks},
    title     = {Physically-Aware Generative Network for 3D Shape Modeling},
    booktitle = {Proceedings of the IEEE/CVF Conference on Computer Vision and Pattern Recognition (CVPR)},
    month     = {June},
    year      = {2021},
    pages     = {9330-9341}
}

@misc{gao2024meshart,
      title={MeshArt: Generating Articulated Meshes with Structure-guided Transformers}, 
      author={Daoyi Gao and Yawar Siddiqui and Lei Li and Angela Dai},
      year={2024},
      eprint={2412.11596},
      archivePrefix={arXiv},
      primaryClass={cs.CV},
      url={https://arxiv.org/abs/2412.11596}, 
}

@inproceedings{yu2022point,
  title={Point-bert: Pre-training 3d point cloud transformers with masked point modeling},
  author={Yu, Xumin and Tang, Lulu and Rao, Yongming and Huang, Tiejun and Zhou, Jie and Lu, Jiwen},
  booktitle={Proceedings of the IEEE/CVF conference on computer vision and pattern recognition},
  pages={19313--19322},
  year={2022}
}

@article{chang2015shapenet,
  title={Shapenet: An information-rich 3d model repository},
  author={Chang, Angel X and Funkhouser, Thomas and Guibas, Leonidas and Hanrahan, Pat and Huang, Qixing and Li, Zimo and Savarese, Silvio and Savva, Manolis and Song, Shuran and Su, Hao and others},
  journal={arXiv preprint arXiv:1512.03012},
  year={2015}
}

@article{vaswani2017attention,
  title={Attention is all you need},
  author={Vaswani, Ashish and Shazeer, Noam and Parmar, Niki and Uszkoreit, Jakob and Jones, Llion and Gomez, Aidan N and Kaiser, {\L}ukasz and Polosukhin, Illia},
  journal={Advances in neural information processing systems},
  volume={30},
  year={2017}
}

@article{loshchilov2017decoupled,
  title={Decoupled weight decay regularization},
  author={Loshchilov, Ilya and Hutter, Frank},
  journal={arXiv preprint arXiv:1711.05101},
  year={2017}
}

@inproceedings{kim2024partstad,
  title={Partstad: 2d-to-3d part segmentation task adaptation},
  author={Kim, Hyunjin and Sung, Minhyuk},
  booktitle={European Conference on Computer Vision},
  pages={422--439},
  year={2024},
  organization={Springer}
}
}
 \clearpage
\setcounter{page}{1}
\maketitlesupplementary

\section{Off-the-shelf segmentation }
We evaluated the robustness of our approach using part segmentations produced by PartSTAD \cite{kim2024partstad}, a few-shot, VLM-based point cloud model. Since VLMs perform surface segmentation, we filter out object classes with interior structures and test on the following classes: \textit{box}, \textit{dishwasher}, \textit{door}, \textit{eyeglasses}, \textit{refrigerator}, and \textit{trash can}.

Table \ref{table:partstad_quant_eval_labeled} shows the average quantitative results across these classes in the unlabeled setting. Although there is a slight degradation in metrics compared to ground-truth (``GT'') segmentation, overall performance remains competitive, indicating that our method can maintain reliable articulation predictions under imperfect segmentation. We illustrate this qualitatively in Figure \ref{fig:rebuttal_real} -- our method predicts the correct rotation for test shape even with segmentation noise. 

We note that shape segmentation is orthogonal to our contributions; our method may benefit further from advances in this space.

\section{Implementation details}
Our network stacks 6 layers of cross-attention and 12 layers of self-attention, each with 8 heads. The decoder consists of lightweight 2-layer MLP blocks whose output dimensionality adapts to the target parameter set (pretraining vs. fine-tuning). We use a batch size of 60 and train on 2 NVIDIA A40 GPUs.

\newlength{\failgemimagewidth}
\setlength{\failgemimagewidth}{0.24\columnwidth} 

\newcolumntype{C}{>{\centering\arraybackslash}p{\failgemimagewidth}}

\begin{figure}[t]
\setlength{\tabcolsep}{1pt}
\begin{tabular}{C C C C}

\includegraphics[width=\failgemimagewidth]{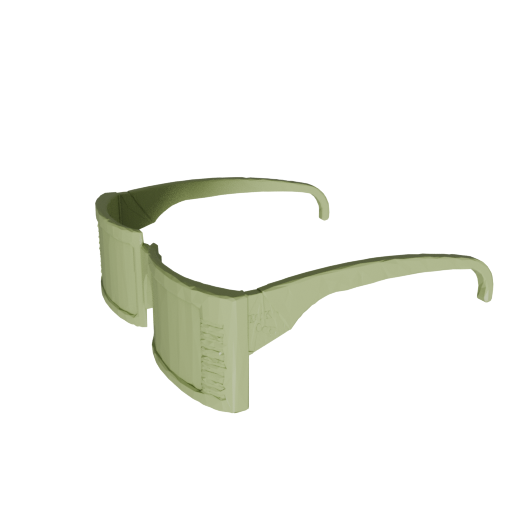} &
\includegraphics[width=\failgemimagewidth]{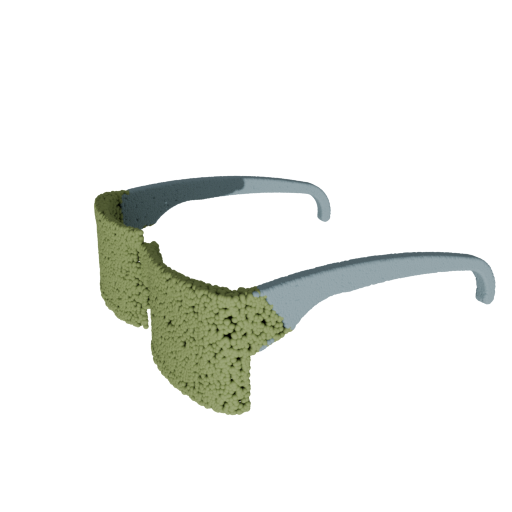} &
\includegraphics[width=\failgemimagewidth]{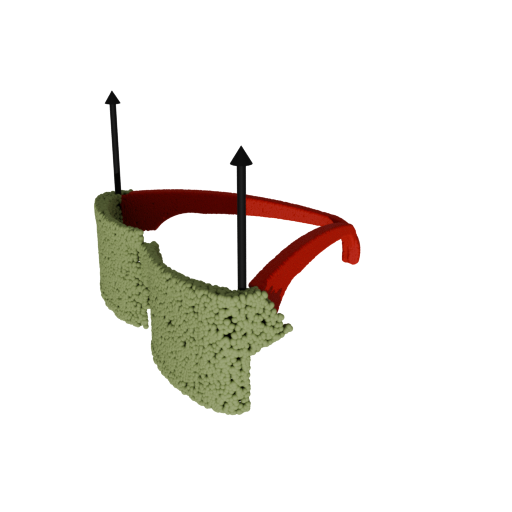} &
\includegraphics[width=\failgemimagewidth]{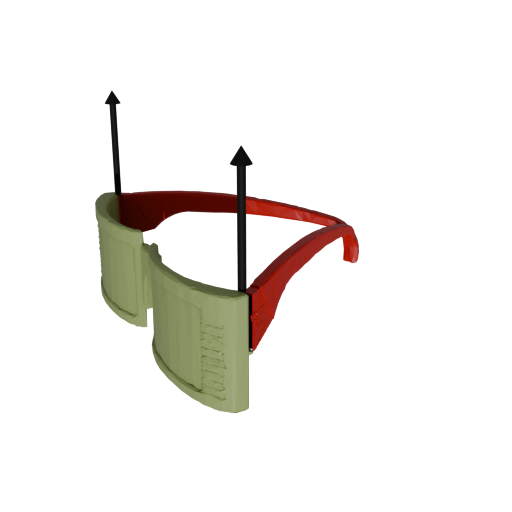}\\

\end{tabular}
\caption{Results using PartSTAD, an off-the-shelf segmentation method. \emph{(1)} input shape, \emph{(2)} PartSTAD segmentation, \emph{(3)} sampled articulated state from predicted parameters.
\emph{(4)} same articulation state from GT segmentation and GT parameters.}
\label{fig:rebuttal_real}
\end{figure}
\begin{table}[t]
\centering
\footnotesize
\begin{tabular}{l|cccc}
\toprule
\textbf{Model} & \textbf{AE $\downarrow$} & \textbf{PE $\downarrow$} & \textbf{R-ACC $\uparrow$} & \textbf{P-ACC $\uparrow$} \\
\midrule
GEOPARD-u-GT  & 7.24& 0.07& 0.96 & 0.98 \\
GEOPARD-u-PartSTAD    &9.34&0.08&0.95&0.93\\

\bottomrule
\end{tabular}
\caption{Using off-the-shelf vs ground-truth segmentation.}
\label{table:partstad_quant_eval_labeled}
\end{table}

\end{document}